\documentclass[12pt]{llncs}
\usepackage{amsfonts}
\usepackage{algorithmicx}
\usepackage[ruled,section]{algorithm}
\usepackage[noend]{algpseudocode}
\usepackage{comment}
\usepackage{graphicx}
\usepackage{url}
\title{An $O(M(n) \log n)$ algorithm for the Jacobi symbol}
\author{Richard P. Brent\inst{1} \and Paul Zimmermann\inst{2}}
\institute{Australian National University, Canberra, Australia
\and
INRIA Nancy - Grand Est, Villers-l\`es-Nancy, France\\[10pt]
28 January 2010\\Submitted to ANTS IX}
\algnewcommand\algorithmicinput{\textbf{Input:}}
\algnewcommand\Input{\item[\algorithmicinput]}%
\algnewcommand\algorithmicoutput{\textbf{Output:}}
\algnewcommand\Output{\item[\algorithmicoutput]}%

\newcommand{\Otilde}{\makebox{$\widetilde O$}}
\newcommand{\ddiv}{\;{\rm div}\;}
\def\N{{\mathbb N}}
\def\Z{{\mathbb Z}}
\renewenvironment{proof}[1]{\trivlist \item[\hskip \labelsep{\em #1}]}{\hfill\mbox{$\fbox{}$}\endtrivlist}
\begin{document}
\maketitle

\begin{abstract}
The best known algorithm to compute the Jacobi symbol of two $n$-bit integers
runs in time $O(M(n)\log n)$, using Sch\"onhage's fast continued fraction
algorithm combined with an identity due to Gauss. We give a different
$O(M(n)\log n)$ algorithm
based on the binary recursive gcd algorithm of 
Stehl\'e and Zimmermann. %
Our implementation --- which to our knowledge is the first to run in 
time $O(M(n)\log n)$ --- is faster than GMP's quadratic implementation for
inputs larger than about $10000$ decimal digits.
\end{abstract}

\section{Introduction} \label{sec:intro}

We want to compute the Jacobi symbol\footnote{Notation: we write the Jacobi symbol
as $(b|a)$, since this is easier to typeset and less ambiguous than
the more usual $\left(\frac{b}{a}\right)$.
$M(n)$ is the time to multiply $n$-bit numbers.
$\Otilde(f(n))$ means $O(f(n)(\log f(n))^c)$ 
for some constant $c \ge 0$.} %
$(b|a)$ for
$n$-bit integers $a$ and $b$, where $a$ is odd positive.
We give three algorithms based on the
$2$-adic gcd from Stehl\'e and Zimmermann~\cite{StZi04}.
First we give an algorithm whose worst-case time bound
is $O(M(n)n^2) = \Otilde(n^3)$;
we call this the {\em cubic} algorithm although this
is pessimistic since the algorithm is quadratic on average as shown in
\cite{DaMaVa05}, and probably also in the worst case.
We then show how to reduce the worst-case to $O(M(n)n) = \Otilde(n^2)$
by combining sequences of ``ugly'' iterations (defined in
Section~\ref{sec:cubic}) into one ``harmless'' iteration.
Finally, we obtain an algorithm with worst-case
time $O(M(n)\log n)$.
This is, up to a constant factor, 
the same as the time bound for the best
known algorithm, apparently never published in full, but
sketched in Bach~\cite{Bach90} and in more detail in 
Bach and Shallit~\cite{BaSh96} (with credit to Bachmann~\cite{Bachmann02}).

The latter algorithm makes use of the Knuth-Sch\"onhage fast
continued fraction algorithm~\cite{Schonhage71b}
and an identity of Gauss~\cite{Gauss65}.
Although this algorithm has been attributed to Sch\"onhage,
Sch\"onhage himself gives a different $O(M(n)\log n)$ 
algorithm~\cite{Schonhage09,Weilert02} which does not depend on the identity
of Gauss. The algorithm is mentioned in Sch\"onhage's
book~\cite[\S7.2.3]{Schonhage94}, but no details are given there. 

With our algorithm it is not necessary to compute the full continued
fraction or to use the identity of Gauss for the Jacobi symbol. Thus, it
provides an alternative that may be easier to implement.

It is possible to modify some of the other fast GCD algorithms
considered by M\"oller~\cite{Moeller08} to compute the Jacobi symbol,
but we do not consider such possibilities here. At best they give a small
constant factor speedup over our algorithm.

We recall the main identities satisfied by the Jacobi symbol:
$(bc|a) = (b|a) (c|a)$;
$(2|a) = (-1)^{(a^2-1)/8}$; $(b|a) = (-1)^{(a-1)(b-1)/4}(a|b)$
for $a, b$ odd;
and $(b|a)=0$ if $(a,b) \ne 1$. %

Note that all our algorithms compute $(b|a)$ with $b$ even positive and
$a$ odd positive. For the more general case where $b$ is any integer,
we can reduce to $b$ even and positive using $(b|a) = (-1)^{(a-1)/2} (-b|a)$
if $b$ is negative, and $(b|a) = (b\!+\!a|a)$ if $b$ is odd.

We first describe a cubic algorithm to compute the Jacobi symbol. The
quadratic algorithm in Section~\ref{sec:harmless-iterations} is 
based on this cubic algorithm, and the subquadratic algorithm
in Section~\ref{sec:sub-quadratic} uses the same ideas as the 
quadratic algorithm but with an asymptotically fast recursive implementation.

For $a \in \Z$, the notation $\nu(a)$ denotes the $2$-adic valuation
$\nu_2(a)$ of 
$a$, that is the maximum $k$ such that $2^k|a$, or $+\infty$ if $a = 0$.

\subsection{Binary Division with Positive Quotient} \label{sec:cubic}

Throughout the paper we use the binary division with positive quotient defined
by Algorithm~\ref{BinaryDividePos}.
Compared to the ``centered division'' of~\cite{StZi04}, it returns a quotient
in $[1, 2^{j+1}-1]$ instead of in $[1-2^j, 2^j-1]$. 
Note that the quotient $q$ is always odd.

\begin{algorithm}[htp]
\caption{BinaryDividePos} \label{BinaryDividePos}
\begin{algorithmic}[1]
\Input $a, b \in \N$ with $\nu(a) = 0 < \nu(b) = j$
\Output $q$ and $r = a + q b/2^j$ such that $0 < q < 2^{j+1}$, $\nu(b) < \nu(r)$
\State $q \leftarrow - a/(b/2^j) \bmod 2^{j+1}$ 
  \Comment{$q$ is odd and positive} \label{BDP:step1}
\State \textbf{return} $q, r = a + q b/2^j$.
\end{algorithmic}
\end{algorithm}

With this binary division, we define Algorithm Cubic\-Binary\-Jacobi,
where the fact that the quotient $q$ is positive
ensures that all 
$a,b$ terms computed remain positive, and $a$ remains odd, thus
$(b|a)$ remains well-defined.\footnote{M\"oller says in
\cite{Moeller08}: ``\emph{if one tries to use positive
quotients $0 < q < 2^{k+1}$, the} [binary gcd] \emph{algorithm no longer
terminates}''. However, with a modified stopping criterion as in
Algorithm CubicBinaryJacobi, the algorithm terminates (we prove this
below).}
\begin{algorithm}[htp]
\caption{CubicBinaryJacobi}
\begin{algorithmic}[1]
\Input $a, b \in \N$ with $\nu(a) = 0 < \nu(b)$
\Output Jacobi symbol $(b|a)$
\State $s \leftarrow 0$, \quad
       $j \leftarrow \nu(b)$
\While{$2^j a \neq b$}
   \State $b' \leftarrow b/2^j$
   \State $(q,r) \leftarrow \mathrm{BinaryDividePos}(a,b)$    \label{CBJ:step5}
   \State $s \leftarrow (s + {j(a^2-1)/8} + {(a-1)(b'-1)/4} + 
		{j({b'}^2-1)/8}) \bmod 2$ 		      \label{CBJ:step6}
   \State $(a, b) \leftarrow (b', r/2^j)$, \quad
          $j \leftarrow \nu(b)$
\EndWhile
\State {\bf if} $a=1$ {\bf then} return $(-1)^s$ {\bf else} return $0$
\end{algorithmic}
\end{algorithm}

\begin{theorem}
Algorithm Cubic\-Binary\-Jacobi is correct (assuming it terminates).
\end{theorem}
\begin{proof}{Proof.}
We prove that the following invariant holds during the algorithm, if
$a_0, b_0$ are the initial values of $a, b$:
\[ (b_0|a_0) = (-1)^s (b|a). \]
This is true before we enter the while-loop, since $s=0$, $a = a_0$, and
$b = b_0$. For each step in the while loop, we divide $b$ by $2^j$,
swap $a$ and $b'=b/2^j$,
replace $a$ by $r = a + q b'$, and divide $r$ by $2^j$.
The Jacobi symbol is modified by a factor
$(-1)^{j(a^2-1)/8}$ for the division of $b$ by $2^j$, by a factor
$(-1)^{(a-1)(b'-1)/4}$ for the interchange of $a$ and $b'$, and by a factor
$(-1)^{j({b'}^2-1)/8}$ for the division of $r$ by $2^j$.
At the end of the loop, we have $\gcd(a_0,b_0)=a$; if $a = 1$, since
$(b|1) = 1$, we have $(b_0|a_0) = (-1)^s$,
otherwise $(b_0|a_0) = 0$.
\end{proof}

\begin{lemma}
The quantity $a + 2b$ is non-increasing in Algorithm Cubic\-Binary\-Jacobi.
\end{lemma}
\begin{proof}{Proof.}
At each iteration of the ``while'' loop, 
$a$ becomes $b/2^j$, and $b$ becomes $(a+qb/2^j)/2^j$.
In matrix notation
\begin{equation}
\left(\begin{array}{c} a \\ b \end{array}\right) \leftarrow
\left(\begin{array}{cc}     0 	& 1/2^j \\
			  1/2^j	& q/2^{2j} \end{array}\right)
\left(\begin{array}{c} a \\ b \end{array}\right).
\label{eq:matrix-iteration}
\end{equation}
Therefore $a+2b$ becomes
\begin{equation} \frac{b}{2^j} + 2\left(\frac{a+qb/2^j}{2^j}\right)
   = \frac{2a}{2^j} + (1+2q/2^j) \frac{b}{2^j}. 	\label{eq:a2b}
\end{equation}
Since $j \geq 1$, the first term is bounded by $a$.
In the second term, $q \leq 2^{j+1}-1$, thus
the second term is bounded by $(5/2^j-2/2^{2j}) b$, which is bounded by
$9b/8$ for $j \geq 2$, and equals $2b$ for $j=1$.
\end{proof}
If $j \geq 2$, then $a+2b$ is multiplied by a factor at most $9/16$.
If $j = q = 1$ then $a+2b$ decreases, but by a factor which could be
arbitrarily close to~$1$.
The only case where $a+2b$ does not decrease is when $j=1$ and $q=3$;
in this case $a+2b$ is unchanged.

This motivates us to define three classes of iterations: 
{\em good}, {\em bad}, and {\em ugly}.
Let us say that we have a {\em good} iteration when $j \geq 2$,
a {\em bad} iteration when $j = q = 1$,
and an {\em ugly} iteration when $j=1$ and $q=3$.
Since $q$ is odd and $1 \leq q \leq 2^{j+1}-1$, 
this covers all possibilities.
For a bad iteration, $(a,b)$ becomes $(b/2, a/2+b/4)$,
and for an ugly iteration, $(a,b)$ becomes $(b/2, a/2+3b/4)$. 
We denote the matrices corresponding to good, bad and ugly iterations
by $G$, $B$ and $U$ respectively.  Thus

\[
G = G_{j,q} = 
\left(\begin{array}{cc}     0 	& 1/2^j \\
			  1/2^j	& q/4^j 
\end{array}\right),\;
B = 
\left(\begin{array}{cc}     0 	& 1/2 \\
			  1/2	& 1/4 
\end{array}\right),\;
U = 
\left(\begin{array}{cc}     0 	& 1/2 \\
			  1/2	& 3/4 
\end{array}\right).
\]
The effect of $m$ successive ugly iterations is easily seen to be 
given by the matrix
\begin{equation}
U^m = \frac{1}{5}
\left(\begin{array}{ll} 1 + 4(-1/4)^m &\; 2 - 2(-1/4)^m \\[2pt]
                        2 - 2(-1/4)^m &\; 4 + (-1/4)^m 
\end{array}
\right).						\label{eq:Um}
\end{equation}
Assume we start from $(a,b) = (a_0,b_0)$,
and after $m>0$ successive ugly iterations we get values $(a_m, b_m)$.
Then, from Equation~(\ref{eq:Um}),
\begin{eqnarray}
  5a_m &=& (a + 2b) + 2(2a-b)(-1/4)^m, 		\label{eq:am} \\
  5b_m &=& 2(a+2b)  - (2a-b)(-1/4)^m.		\label{eq:bm}
\end{eqnarray}
We can not have $2a_0 = b_0$ or the algorithm would have terminated.
However, $a_m$ must be an integer. This gives an upper bound on $m$.
For $a_0,b_0$ of $n$ bits, the number of successive ugly iterations is bounded
by $n/2 + O(1)$ (a precise statement is made in Lemma~\ref{lemma2}).

If there were no bad iterations, this would prove that for $n$-bit
inputs the number of iterations is $O(n^2)$, since each sequence of ugly
iterations would be followed by at least one good iteration.  
Bad iterations
can be handled by a more complicated argument which we omit, since they
will be considered in detail in \S\ref{sec:harmless-iterations} 
when we discuss the complexity of the quadratic algorithm
(see the proof of Theorem~\ref{thm:qbj}).

Since the number of iterations is $O(n^2)$ from Theorem~\ref{thm:qbj},
and each iteration costs time $O(M(n))$, the overall time for Algorithm
Cubic\-Binary\-Jacobi is $O(n^2 M(n)) = \Otilde(n^3)$.
Note that this worst-case bound is almost certainly too pessimistic
(see \textsection\ref{results}).

\section{A Provably Quadratic Algorithm}
\label{sec:harmless-iterations}

Suppose we have a sequence of $m > 0$ ugly iterations.
It is possible to combine the $m$ ugly iterations into
one {\em harmless} iteration which is not much more expensive than a
normal (good or bad) iteration. 
Also, it is possible to predict the maximal such $m$ in advance. 
Using this trick, we reduce the number of iterations 
(good, bad and harmless) to
$O(n)$ and their cost to $O(M(n)n) = \Otilde(n^2)$.

Without loss of generality, suppose that we start from $(a_0,b_0)$ =
$(a,b)$. Since $a$ is odd, we never have $a=2b$. 

\begin{lemma} \label{lemma2}
If $\mu = \nu(a-b/2)$, then we have exactly $\lfloor \mu/2 \rfloor$
ugly iterations starting from $(a, b)$, followed by a good iteration if
$\mu$ is even, and by a bad iteration if $\mu$ is odd.
\end{lemma}
\begin{proof}{Proof.}
We prove the lemma by induction on $\mu$.
If $\mu=0$, $a-b/2$ is odd, but $a$ is odd, so $b/2$ is even,
which yields $j \geq 2$ in Binary\-Divide\-Pos, 
thus $a,b$ yield a good iteration.
If $\mu=1$, $a-b/2$ is even, which implies that $b/2$ is odd,
thus we have $j=1$. If we had $q=3$ in Binary\-Divide\-Pos, this would
mean that $a + 3 (b/2) = 0 \bmod 4$, or equivalently
$a - b/2 = 0 \bmod 4$, which is incompatible with $\mu=1$.
Thus we have $q=1$, and a bad iteration.

Now assume $\mu \geq 2$. 
The first iteration is ugly since $4$ divides
$a-b/2$, which implies that $b/2$ is odd. Thus $j=1$, and
\hbox{$a - b/2 = 0 \bmod 4$} implies that $q=3$.
After one ugly iteration $(a,b)$ becomes
$(b/2,a/2+3b/4)$, thus $a-b/2$ becomes $-(a-b/2)/4$,
and the $2$-valuation of $a-b/2$ decreases by $2$.
\end{proof}

From the above, we see that,
for a sequence of $m$ ugly iterations,
$a_0,a_1,\ldots,a_m$ satisfy the three-term
recurrence
\[
4a_{i+1} - 3a_i - a_{i-1} = 0 \;\; {\rm for} \;\; 0 < i < m,
\]
and similarly for $b_0,b_1,\ldots,b_m$.
It follows that $a_i = a \bmod 4$, 
and similarly $b_i = b \bmod 4$, for $1 \le i < m$.

We can modify Algorithm Cubic\-Binary\-Jacobi to consolidate $m$ 
consecutive ugly
iterations into one harmless iteration, 
using the expressions (\ref{eq:am})--(\ref{eq:bm}) for
$a_m$ and $b_m$ (we give an optimised evaluation below).  It remains
to modify step~\ref{CBJ:step6} of CubicBinaryJacobi
to take account of the $m$ updates to $s$. Since $j=1$ for
each ugly iteration, we have to increment $s$ by an amount 
\[
\delta = \sum_{0 \le i < m} \left( \frac{a_i^2-1}{8} + 
			\frac{{b'_i}^2-1}{8} + 
			\frac{a_i-1}{2} \frac{b'_i-1}{2} \right) \bmod 2,
\]
where we write $b'_i$ for $b_i/2$.
However, $a_{i+1} = b'_i$ for $0 \le i < m$, so the terms involving
division by $8$ ``collapse'' mod~$2$, leaving just the first and last terms.
The terms involving two divisions by $2$ are all equal to
$(a-1)/2 \cdot (b'-1)/2$
mod~$2$, using the observation that $a_i$ mod~$4$ is
constant for $0 \le i \le m$. 
Thus
\[
\delta = \left( \frac{a_0^2-1}{8} + \frac{a_m^2-1}{8} +
	m\frac{a_0-1}{2} \frac{a_1-1}{2} \right) \bmod 2.
\]
One further simplification is possible. Since $a_0 = a_1 \bmod 4$,
and $a_0$ is odd, we can
replace $a_1$ by $a_0$ in the last term, and use the fact that
$x^2 = x \bmod 2$ to obtain
\begin{equation}
\delta = \left( \frac{a_0^2-1}{8} + \frac{a_m^2-1}{8} +
	m\frac{a_0-1}{2} \right) \bmod 2.		\label{eq:delta}
\end{equation}
We can economise the computation of $a_m$ and $b_m$ 
from (\ref{eq:am})--(\ref{eq:bm}) by first computing
\[d = a - b',\; m = \nu(d) \ddiv 2,\; c = (d - (-1)^m(d/4^m))/5,\]
where the divisions by $4^m$ and by $5$ are exact; then
$a_m = a - 4c$, $b_m = b + 2c$.

From these observations, it is easy to modify Algorithm 
Cubic\-Binary\-Jacobi to
obtain Algorithm Quadratic\-Binary\-Jacobi. In this algorithm, 
steps \ref{QBJ:step8}--\ref{QBJ:step12}
implement a harmless iteration equivalent to $m > 0$ consecutive ugly
iterations; steps \ref{QBJ:step14}--\ref{QBJ:step15} 
implement bad and good iterations,
and the remaining steps are common to both. 
Step~\ref{CBJ:step6} of Algorithm Cubic\-Binary\-Jacobi 
is split into three steps
\ref{QBJ:step5}, \ref{QBJ:step14} and \ref{QBJ:step16}.
In the case of a harmless iteration, the computation of
$\delta$ satisfying~(\ref{eq:delta}) is implicit in steps 
\ref{QBJ:step5}, \ref{QBJ:step11} and \ref{QBJ:step16}.

\begin{algorithm}[htp]
\caption{QuadraticBinaryJacobi}
\begin{algorithmic}[1]
\Input $a, b \in \N$ with $\nu(a) = 0 < \nu(b)$
\Output Jacobi symbol $(b|a)$
\State $s \leftarrow 0$, \quad
       $j \leftarrow \nu(b)$
\While{$2^j a \neq b$}					\label{QBJ:step3}
   \State $b' \leftarrow b/2^j$
   \State $s \leftarrow (s + j(a^2-1)/8) \bmod 2$ 	\label{QBJ:step5}
   \State $(q,r) \leftarrow \mathrm{BinaryDividePos}(a,b)$
   \If{$(j,q) = (1,3)$}
      \State $d \leftarrow a - b'$ 			\label{QBJ:step8}
      \State $m \leftarrow \nu(d) \ddiv 2$
      \State $c \leftarrow (d - (-1)^m d/4^m)/5$
      \State $s \leftarrow (s + m(a-1)/2) \bmod 2$ 	\label{QBJ:step11}
      \State $(a, b) \leftarrow (a - 4c, b + 2c)$ 
			\Comment{harmless iteration}  	\label{QBJ:step12}
   \Else
     \State $s \leftarrow (s + (a-1)(b'-1)/4) \bmod 2$ 	\label{QBJ:step14}
     \State $(a, b) \leftarrow (b', r/2^j)$ 		
			\Comment{good or bad iteration}	\label{QBJ:step15}
   \EndIf
   \State $s \leftarrow (s + j(a^2-1)/8) \bmod 2$, 	\label{QBJ:step16}
          \quad
          $j \leftarrow \nu(b)$				\label{QBJ:step17}
\EndWhile
\State {\bf if} $a=1$ {\bf then} return $(-1)^s$ {\bf else} return $0$
\end{algorithmic}
\end{algorithm}

\begin{theorem} 					\label{thm:qbj}
Algorithm Quadratic\-Binary\-Jacobi is correct and terminates after $O(n)$
iterations of the ``while'' loop (steps \ref{QBJ:step3}--\ref{QBJ:step17})
if the inputs are
positive integers of at most $n$ bits, with $0 = \nu(a) < \nu(b)$.
\end{theorem}
\begin{proof}{Proof.}
Correctness
follows from the equivalence to 
Algorithm Cubic\-Binary\-Jacobi. To prove that convergence takes $O(n)$
iterations, we show that $a+2b$ is multiplied by a 
factor at most $5/8$ in each block of three iterations. This is true if
the block includes at least one good iteration, so we need only consider
harmless and bad iterations. Two harmless iterations do not occur in
succession, so the block must include either (harmless, bad) or
(bad, bad).  In the first case, the corresponding matrix is 
$BU^m = BU\cdot U^{m-1}$ for some $m>0$.
We saw in \S\ref{sec:cubic} that the matrix $U$ leaves $a+2b$ 
unchanged,
so $U^{m-1}$ also leaves $a+2b$ unchanged, and we need only consider the 
effect of $BU$.
Suppose that $(a, b)$ is transformed into $(\widetilde{a},\widetilde{b})$
by $BU$. Thus
\[
\left(\begin{array}{c} \widetilde{a} \\ \widetilde{b} \end{array}\right) =
BU
\left(\begin{array}{c} {a} \\ {b} \end{array}\right) =
\left(\begin{array}{c@{\quad}c}	1/4	& 3/8 \\ 1/8 & 7/16
\end{array}\right)
\left(\begin{array}{c} {a} \\ {b} \end{array}\right)\;.
\]
We see that
\[\widetilde{a} + 2\widetilde{b} = \frac{a}{2} + \frac{5b}{4} \le 
	\frac{5}{8}(a + 2b).
\]
The case of two successive bad iterations is similar~-- just replace $BU$
by $B^2$ in the above, and deduce that 
$\widetilde{a} + 2\widetilde{b} \le (a+2b)/2$.

We conclude that the number of iterations of the while loop is at most
$cn + O(1)$, where $c = 3/\log_2(8/5) \approx 4.4243$.
\end{proof}

\noindent{\em Remarks}\\[5pt]
1. A more complicated argument along similar lines can reduce 
the constant $c$ to 
$2/\log_2(1/\rho(BU)) = 2/\log_2((11-\sqrt{57})/2) \approx 2.5424$.
Here $\rho$ denotes the spectral radius:
$\rho(A) = \lim_{k\to\infty}||A^k||^{1/k}$.\\
2. In practice Quadratic\-Binary\-Jacobi is not much (if any)
faster than Cubic\-Binary\-Jacobi. Its advantage is simply the
better worst-case time bound. A heuristic argument suggests that on average
only $1/4$ of the iterations of Cubic\-Binary\-Jacobi are ugly.\\
3. Our implementations of Cubic\-Binary\-Jacobi and Quadratic\-Binary\-Jacobi
are slower than GMP's $O(n^2)$ algorithm 
(which is based
on Stein's binary gcd, as in Shallit and Sorenson~\cite{ShSo93}). 
However, in the next section we use the ideas
of our Quadratic\-Binary\-Jacobi algorithm to get an $O(M(n)\log n)$
algorithm.  We do not see how to modify the algorithm of Shallit and 
Sorenson to do this.\footnote{In Algorithm
Binary Jacobi in~\cite{ShSo93}, it is necessary to know the
sign of $a-n$ ($b-a$ in our notation) to decide whether to perform
an interchange. This makes it difficult to construct a recursive
$O(M(n)\log n)$ algorithm like Algorithm Half\-Binary\-Jacobi.}

\section{An $O(M(n) \log n)$ Algorithm} \label{sec:sub-quadratic}

Algorithm Half\-Binary\-Jacobi is a modification of Algorithm Half-GB-gcd from
\cite{StZi04}. The main differences are the following:
\begin{enumerate}
\item binary division with positive (not centered) quotient is used;
\item the algorithm returns an integer $s$ such that if $a, b$ are the
      inputs, $c, d$ the output values defined by Theorem~\ref{theorem2}, then
      \[ (b|a) = (-1)^s (d|c); \]
\item at steps~\ref{HBGCD:step4} and~\ref{HBGCD:step27}, we
      reduce mod $2^{2k_1+2}$ (resp.~$2^{2k_2+2}$) 
      instead of mod $2^{2k_1+1}$ (resp.~$2^{2k_2+1}$),
      so that we have enough information to correctly update $s_0$
      at steps \ref{HBGCD:step10}, \ref{HBGCD:step17}, \ref{HBGCD:step21} 
      and \ref{HBGCD:step25};
\item we have to ``cut'' some harmless iterations in two 
      (step~\ref{HBGCD:stepm}).
\end{enumerate}

\begin{algorithm}[htp]
\caption{HalfBinaryJacobi}
\begin{algorithmic}[1]
\Input $a \in \N, b \in \N \cup \{0\}$ with $0 = \nu(a) < \nu(b)$, and $k \in \N$
\Output two integers $s, j$ and a $2 \times 2$ matrix $R$
\If {$\nu(b) > k$}		\Comment{$b = 0$ is possible}
   \State Return $0, 0, 
     \left(\begin{array}{cc}1&0\\0&1\end{array}\right)$	   \label{HBGCD:step2}
\EndIf
\State $k_1 \leftarrow \lfloor k/2 \rfloor$
\State $a_1 \leftarrow a \bmod 2^{2k_1+2}$, \quad
       $b_1 \leftarrow b \bmod 2^{2k_1+2}$ 		   \label{HBGCD:step4}
\State $s_1, j_1, R \leftarrow 
	  {\rm HalfBinaryJacobi}(a_1, b_1, k_1)$ 	   \label{HBGCD:step5}
\State $a' \leftarrow 2^{-2j_1} (R_{1,1} a + R_{1,2} b)$,  \quad
       $b' \leftarrow 2^{-2j_1} (R_{2,1} a + R_{2,2} b)$   \label{HBGCD:step6}
\State $j_0 \leftarrow \nu(b')$
\If {$j_0 + j_1 > k$}
   \State Return $s_1, j_1, R$ \label{HBGCD:step9}
\EndIf
\State $s_0 \leftarrow j_0({a'}^2-1)/8 \bmod 2$		   \label{HBGCD:step10}
\State $q, r \leftarrow {\rm BinaryDividePos}(a', b')$
\State $b'' \leftarrow b'/2^{j_0}$			   \label{HBGCD:step12}
\If{$(j_0,q) = (1,3)$}					   \label{HBGCD:step13}
    \State $d \leftarrow a' - b''$
    \State $m \leftarrow {\rm min}(\nu(d) \ddiv 2, k-j_1)$ \label{HBGCD:stepm}
    \State $c \leftarrow (d - (-1)^m d/4^m)/5$
    \State $s_0 \leftarrow s_0 + m (a'-1)/2 \bmod 2$ 	   \label{HBGCD:step17}
    \State $(a_2, b_2) \leftarrow (a'-4c,2(b''+c))$
		\Comment{harmless iteration}		   \label{HBGCD:stepa2}
    \State $Q \leftarrow \left( \begin{array}{cc} (4^m+4(-1)^m)/5 &
           2(4^m-(-1)^m)/5 \\ 2(4^m-(-1)^m)/5 & (4^{m+1}+(-1)^m)/5 \end{array}
           \right)$ \label{HBGCD:step19}
\Else							   \label{HBGCD:step20}
    \State $s_0 \leftarrow s_0 + (a'-1)(b''-1)/4 \bmod 2$  \label{HBGCD:step21}
    \State $(a_2, b_2) \leftarrow (b'', r/2^{j_0})$
		\Comment{good or bad iteration}
    \State $Q \leftarrow 
      \left(\begin{array}{cc}0&2^{j_0}\\2^{j_0}&q\end{array}\right)$ 
							   \label{HBGCD:step23}
    \State $m \leftarrow j_0$
\EndIf
\State $s_0 \leftarrow s_0 + j_0(a_2^2-1)/8 \bmod 2$	   \label{HBGCD:step25}
\State $k_2 \leftarrow k - (m + j_1)$ 			   \label{HBGCD:step11}
\State $s_2, j_2, S \leftarrow {\rm HalfBinaryJacobi}(a_2 \bmod 2^{2k_2+2}, 
          b_2 \bmod 2^{2k_2+2}, k_2)$ 			   \label{HBGCD:step27}
\State Return $(s_0 + s_1 + s_2) \bmod 2,\, j_1 + j_2 + m$,\, $S \times Q \times R$
\end{algorithmic}
\end{algorithm}

\paragraph{Remarks.} The matrix $Q$ occurring at 
step~\ref{HBGCD:step19} is just $2^{2m} U^m$, where $U^{m}$ is given
by Equation~(\ref{eq:Um}).
Similarly, the matrix $Q$ occurring at 
step~\ref{HBGCD:step23} is $2^{2j_0} G_{j_0,q}$. 
In practice, steps \ref{HBGCD:step13}--\ref{HBGCD:step20} can be omitted
(so the algorithm becomes a fast version of Cubic\-Binary\-Jacobi)~-- this
variant is simpler and slightly faster on average.

\begin{theorem} \label{theorem2}
Let $a, b, k$ be the inputs of Algorithm Half\-Binary\-Jacobi, and
$s, j, R$ the corresponding outputs.
If $\left( c \atop d \right) = 2^{-2j} R \left( a \atop b \right)$, then:
\[ (b|a) = (-1)^s (d|c) \quad \mbox{and} \quad
   \nu(2^j c) \leq k < \nu(2^j d). \]
\end{theorem}
\begin{proof}{Proof (outline).}

We prove the theorem by induction on the parameter $k$.
The key ingredient is that if we reduce $a, b$ mod $2^{2k_1+1}$ in
step~\ref{HBGCD:step4}, then the GB sequence of $a_1, b_1$ matches
that of $a, b$, for the terms computed by the recursive call at
step~\ref{HBGCD:step5}. This is a consequence of~\cite[Lemma~7]{StZi04} 
(which also holds for binary division with positive quotient).
It follows that in all the binary divisions with inputs $a_i, b_i$ in that
recursive call, $a_i$ and $b_i/2^{j_i}$ match modulo $2^{j_i+1}$
the corresponding 
values that would be obtained from the full inputs $a, b$
(otherwise the corresponding binary quotient $q_i$ would
be wrong).
Since here we reduce $a, b$ mod $2^{2k_1+2}$ instead of mod $2^{2k_1+1}$,
$a_i$ and $b_i/2^{j_i}$ now match modulo $2^{j_i+2}$ 
--- instead of modulo $2^{j_i+1}$ ---
the values that would be obtained from the full inputs $a, b$,
where $2^{j_i+2} \geq 8$ since $j_i \geq 1$.

At step~\ref{HBGCD:step10}, $s_0$ depends only on $j_0 \bmod 2$ and
$a' \bmod 8$, at step~\ref{HBGCD:step17} it depends
on $m \bmod 2$ and $a' \bmod 4$,
and at step~\ref{HBGCD:step21} on $a' \bmod 4$ and $b'' \bmod 4$.
Since $a'$ and $b''$ at step~\ref{HBGCD:step21} correspond to some
$a_i$ and $b_i/2^{j_i}$, it follows that $a'$ and $b''$ agree
mod $8$ with the values that would be computed from the full inputs,
and thus the correction $s_0$ is correct. This proves by induction that
$(b|a) = (-1)^s (d|c)$.

Now we prove that $\nu(2^j c) \leq k < \nu(2^j d)$. 
If there is no 
harmless iteration, $\nu(2^j c) \leq k < \nu(2^j d)$
is a consequence of the proof of Theorem~1 in
\cite{StZi04}. In case there is a harmless iteration, first assume that
$m = \nu(d) \ddiv 2$ at step~\ref{HBGCD:stepm}. The new values $a_2,b_2$
at step~\ref{HBGCD:stepa2} correspond to $m$ successive ugly iterations,
which yield $j = j_1 + m \leq k$. Thus $\nu(2^j a_2) \leq k$: we did not
go too far, and since we are computing the same sequence of quotients as
Algorithm QuadraticBinaryJacobi, the result follows.
Now if $k - j_1 < \nu(d) \ddiv 2$, we would go too far if we performed
$\nu(d) \ddiv 2$ ugly iterations, since it would give $j_0 := \nu(d) \ddiv 2
> k - j_1$, thus $j := j_1 + j_0 > k$, and $\nu(2^j a_2)$ would exceed $k$.
This is the reason why we ``cut'' the harmless iteration at $m=k-j_1$ 
(step~\ref{HBGCD:stepm}). The other invariants are unchanged.
\end{proof}

Finally we can present our $O(M(n)\log n)$ Algorithm
Fast\-Binary\-Jacobi, which computes the Jacobi symbol by calling
Algorithm HalfBinaryJacobi.  The general structure is similar to that
described in~\cite{Moeller08} for several asymptotically fast GCD algorithms.

\begin{algorithm}[htp]
\caption{FastBinaryJacobi}
\begin{algorithmic}[1]
\Input $a, b \in \N$ with $0 = \nu(a) < \nu(b)$
\Output Jacobi symbol $(b|a)$
\State $s \leftarrow 0$, \quad
       $j \leftarrow \nu(b)$
\While{$2^j a \ne b$} 
   \State $k \leftarrow \max (\nu(b),\; {\ell}(b) \;{\rm div}\; 3)$
	\Comment{${\ell}(b)$ is length of $b$ in bits}
   \State $s', j, R \leftarrow {\rm HalfBinaryJacobi}(a, b, k)$
   \State $s \leftarrow (s + s') \bmod 2$
   \State $(a, b) \leftarrow 2^{-2j} (R_{1,1}a + R_{1,2}b, R_{2,1}a + R_{2,2}b)$, \quad
          $j \leftarrow \nu(b)$
\EndWhile
\State {\bf if} $a=1$ {\bf then} return $(-1)^s$ {\bf else} return $0$
\end{algorithmic}
\end{algorithm}

Daireaux, Maume-Deschamps and Vall\'ee~\cite{DaMaVa05} prove that,
for the positive binary division,
the average increase of the most significant bits is $0.65$ bits/iteration 
(which partly cancels an average decrease of
two least significant bits per iteration);
compare this with only $0.05$ bits/iteration on average 
for the centered division.\footnote{We have computed more accurate values
of these constants: $0.651993$ and $0.048857$ respectively.}

\section{Experimental Results} \label{results}

We have implemented the different algorithms in C (using $64$-bit integers)
and in GMP (using multiple-precision integers), as well as in
Maple/Magma (for testing purposes).

For $\max(a, b) < 2^{26}$ the maximum number of iterations of Algorithm
Cubic\-Binary\-Jacobi is $64$,
with $a=15548029$ and $b=66067306$.
The number of iterations seems to be
$O(n)$ for $a, b < 2^n$: see Table~1.
This is plausible because, from heuristic probabilistic
arguments, we expect about half of the iterations to be good, and 
experiments confirm this.
For example, if we consider all admissible 
$a, b < 2^{20}$, the cumulated number of
iterations is $3.585\times 10^{12}$ for $2^{38}$ calls, 
i.e., an average of $13.04$ iterations per call (max $48$);
the cumulated number of good, bad and ugly iterations is
$51.78$\%, $25.47$\%, and $22.75$\% respectively.
For $a, b < 2^{60}$, a random
sample of $10^8$ pairs $(a,b)$ gave $42.72$ iterations per call (max $89$),
with $50.54\%$, $25.14$\%, and $24.31\%$ for good, bad and ugly
respectively.  These ratios seem to be converging to the heuristically expected
$1/2=50\%$, $1/4=25\%$, and $1/4=25\%$.

When we consider all admissible $a, b < 2^{20}$, the maximum number of
iterations of Quadratic\-Binary\-Jacobi is $37$ when $a=933531$, $b=869894$,
the cumulated number of iterations is $3.405 \times 10^{12}$ ($12.39$ per call),
the cumulated number of good, bad and harmless iterations is
$54.51$\%, $26.82$\%, and $18.67$\% respectively.
For $a, b < 2^{60}$, a random
sample of $10^8$ pairs $(a,b)$ gave $40.21$ iterations per call (max $76$),
with $53.70\%$, $26.71$\%, and $19.59\%$ for good, bad and harmless
respectively.  These ratios seem to be converging to the heuristically expected
$8/15=53.33\%$, $4/15=26.67\%$, and $1/5 = 20\%$.

\begin{figure}[hb] %
\centerline{\includegraphics[width=10cm]{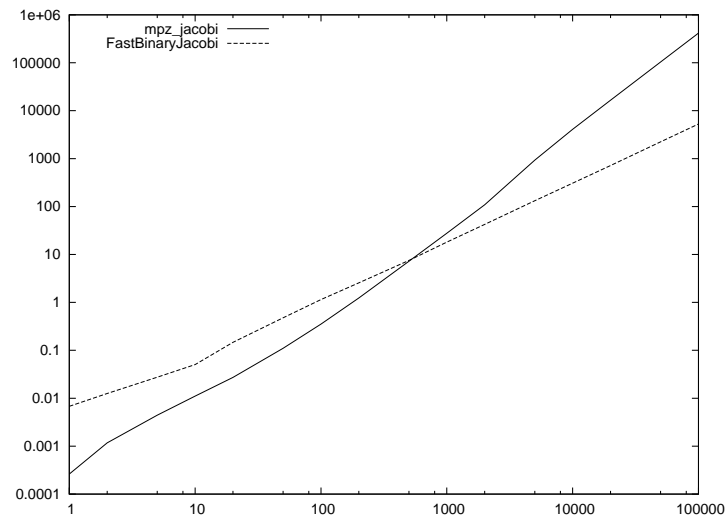}} %
\caption{Comparison of GMP 4.3.1 \texttt{mpz\_jacobi} routine
with our \texttt{FastBinaryJacobi} implementation in log-log scale.
The $x$-axis is in
$64$-bit words, the $y$-axis in milliseconds on a 2.83Ghz Core 2.}
\label{fig1}
\end{figure}

We have also compared the time and average number of iterations for huge
numbers, using the fast gcd algorithm in GMP, say \texttt{gcd}
--- which implements
the algorithm from \cite{Moeller08} --- and an implementation of the
algorithm from \cite{StZi04}, say \texttt{bgcd}.
For inputs of one million $64$-bit words, \texttt{gcd} takes about 45.8s on
a 2.83Ghz Core 2, while \texttt{bgcd} takes about 48.3s and 32,800,000
iterations: this is in accordance with the fact proven in \cite{DaMaVa05} that
each step of the binary gcd discards on average two least significant bits,
and adds on average about 0.05 most significant bits.
Our algorithm \texttt{bjacobi}
(based on Algorithms 3.1--3.2)
takes about 83.1s and 47,500,000 iterations (for a version with
steps \ref{HBGCD:step13}--\ref{HBGCD:step20} of Algorithm 3.1 omitted
in the basecase routine), 
which agrees with the theoretical drift of $0.651993$ bits per iteration.
The break-even point between the $O(n^2)$ implementation of the Jacobi symbol
in GMP 4.3.1 and our $O(M(n) \log n)$ implementation is about 
$535$ words, that is about $34,240$ bits or
about $10,300$ decimal digits (see Fig.~\ref{fig1}).
\vspace*{-7pt}

\section{Concluding Remarks}
\vspace*{-3pt}

Weilert~\cite{Weilert02} says: ``\emph{We are not able to use a GCD
calculation in $\Z[i]$ similar to the binary GCD algorithm $\cdots$ 
because we
do not get a corresponding quotient sequence in an obvious manner}''. In a
sense we filled that gap for the computation of the Jacobi symbol, because
we showed how it can be computed using a binary GCD algorithm without the
need for a quotient sequence.

We showed how to compute the Jacobi symbol with an asymptotically fast
time bound, using a binary GCD algorithm without the need for a quotient 
sequence. Our implementation is faster than a good $O(n^2)$ implementation
for numbers with bitsize $n > 35000$.
Our subquadratic implementation %
is available from \url{http://www.loria.fr/~zimmerma/software/#jacobi}.

Binary division with a centered quotient does not seem to give a
subquadratic algorithm; however we can use it with the ``cubic'' algorithm
(which then becomes provably quadratic)
since then we control the sign of $a, b$.
For a better quadratic algorithm,
we can choose the quotient $q$ so that $abq < 0$,
by replacing $q$ by $q-2^{j+1}$ if necessary:\linebreak[3]
experimentally, this
gains on average $2.194231$ bits per iteration, 
compared to $1.951143$ for the
centered quotient, and $1.348008$ for the positive quotient.
In comparison, Stein's ``binary'' algorithm gains on average $1.416488$ bits
per iteration
\cite[\textsection 7]{rpb183}\cite[\textsection 4.5.2]{Knuth97}.
\begin{table}[hb]
\begin{center}
\begin{tabular}{|c|c|c||c|c|c|}
\hline
$n$		& {\rm iterations} 	& example $(a,b)$&
$n$		& {\rm iterations} 	& example $(a,b)$\\
\hline
5  & 6  & $(7, 30)$&
22 & 53 & $(2214985,2781506)$\\
10 & 19 & $(549, 802)$&
23 & 55 & $(1383497,8292658)$\\
15 & 34 & $(23449, 19250)$&
24 & 58 & $(2236963,12862534)$\\
20 & 48 & $(656227, 352966)$&
25 & 62 & $(28662247, 30847950)$\\
21 & 51 & $(1596811,1493782)$&
26 & 64 & $(15548029,66067306)$\\
\hline
\end{tabular}    \label{tab:CBJtable} %
\vspace*{5pt}
\caption{Worst cases for CubicBinaryJacobi$(b|a)$, $\max(a,b) < 2^n$.}
\end{center}
\end{table}		
\paragraph{Acknowledgement.} The authors thank Steven Galbraith who asked them
about the existence of an $O(M(n) \log n)$ algorithm for the\linebreak 
Jacobi symbol, 
Arnold Sch\"onhage for his comments
and a pointer to the work of his former student Andr\'e Weilert, 
Damien Stehl\'e who suggested adapting the binary gcd algorithm,
and Marco Bodrato and Niels M\"oller for testing our implementation.
We also thank INRIA for its support of the ANC ``\'equipe associ\'ee''.
The first author acknowledges the support of the Australian Research Council.

\end{document}